\documentclass[aps,pra,floatfix,a4]{revtex4-1}
\usepackage{graphicx}
\usepackage[hypertexnames=false]{hyperref}
\hypersetup{
	bookmarks=true,         
	unicode=false,          
	pdftoolbar=true,        
	pdfmenubar=true,        
	pdffitwindow=false,     
	pdfstartview={FitH},    
	pdftitle={My title},    
	pdfauthor={Author},     
	pdfsubject={Subject},   
	pdfcreator={Creator},   
	pdfproducer={Producer}, 
	pdfkeywords={keyword1, key2, key3}, 
	pdfnewwindow=true,      
	colorlinks=true,       
	linkcolor=blue,          
	citecolor=blue,        
	filecolor=blue,         
	urlcolor=blue        
}

\usepackage{color}

\usepackage{fancyhdr}

\usepackage{amssymb,amsfonts,amsmath}
\usepackage{xurl}

\newcommand{\GR}{G_{\rm R}}

\newcommand{\sign}{\mathrm{sign}}

\begin{document} 

\title {COVID-19 impact on the international trade}

        \author{C\'elestin Coquid\'e, Jos\'e Lages*}
        \affiliation{Institut UTINAM, OSU THETA, 
            Universit\'e Bourgogne Franche-Comt\'e, CNRS, Besançon, France}
	\author{Leonardo Ermann}
	\affiliation{Departamento de F\'{\i}sica Te\'orica, GIyA,
         Comisi\'on Nacional de Energ\'{\i}a At\'omica.
           Av.~del Libertador 8250, 1429 Buenos Aires, Argentina}
         \affiliation{Consejo Nacional de Investigaciones
             Cient\'ificas y T\'ecnicas (CONICET), C1425FQB, Buenos Aires, Argentina}
	\author{Dima L. Shepelyansky}
	\affiliation{\mbox{Laboratoire de Physique Th\'eorique, IRSAMC, 
	Universit\'e de Toulouse, CNRS, UPS, 31062 Toulouse, France}}

\date{January 18, 2022}


\begin{abstract}
Using the United Nations Comtrade database, we perform the Google matrix analysis of the multiproduct World Trade Network (WTN) for the years 2018-2020 comprising the emergence of the COVID-19 as a global pandemic. The applied algorithms -- the PageRank, the CheiRank and the reduced Google matrix -- take into account the multiplicity of the WTN links providing new insights on the international trade comparing to the usual import-export analysis. These algorithms establish new rankings and trade balances of countries and products considering every countries on equal grounds, independently of their wealth, and every products on the basis of their relative exchanged volumes. In comparison with the pre-COVID-19 period, significant changes in these metrics occur for the year 2020 highlighting a major rewiring of the international trade flows induced by the COVID-19 pandemic crisis. We define a new PageRank-CheiRank product trade balance, either export or import oriented, which is significantly perturbed by the pandemic.
\end{abstract}

\maketitle

\section{Introduction}
The COVID-19 pandemic produced an enormous impact
on the human society with multiple health, social and economic effects
\cite{covidwiki}.
The impact on the international trade is enormous. Thus the World Trade Organization (WTO)
states that ``the COVID-19 pandemic represents an unprecedented disruption 
to the global economy and world trade, as production and consumption 
are scaled back across the globe'' \cite{wto1}. 
Several reports of the WTO and the United Nations (UN) highlight the effects of the COVID-19
for the world trade \cite{wto2,un1,un2}. 
A need for an effective public health response to the collapse of global trade
is elucidated in \cite{lancet}.

The WTO and the UN reports \cite{wto1,wto2,un1,un2} depict the global change
of the world trade induced by the  COVID-19. However, it is also important
to analyze how the trade relations between the countries are affected by the pandemic.
With this aim, we perform the Google matrix analysis of the World Trade Network (WTN)
constructed from the UN Comtrade database \cite{comtrade} for the years 2018, 2019, and 2020. The WTN concerns $N_p=10$ types of commercial products (see Tab.~\ref{tab1})  traded between $N_c = 194$ world countries.
After the pioneering work of Brin and Page \cite{brin}, 
the invented PageRank algorithm, based implicitly on the Google matrix construction, found broad
applications for the World Wide Web (WWW) search engines \cite{brin,meyer}.
In the last decades, the human society created a variety of complex networks
which the main properties are described, for example, in \cite{dorogovtsev}.
The applications of Google matrix algorithms to various directed networks include:
university web sites and 
Wikipedia networks \cite{wikizzs,rmp2015}, Linux Kernel networks \cite{linux1,linux2},
networks of protein-protein interactions \cite{proteinplos,signor,metacore} and many others.

\begin{table}[!h]\centering
	\begin{tabular}{|c|l|}
		\hline
		SITC code & Standard International Trade Classification (SITC) product groups\\
		\hline\hline
		0 & Food and live animals\\
		1 & Beverages and tobacco\\
		2 & Crude materials, inedible, except fuels\\
		3 & Mineral fuels, lubricants and related materials\\
		4 & Animal and vegetable oils, fats and waxes\\
		5 & Chemicals and related products, n.e.s.\\
		6 & Manufactured goods classified chiefly by material\\
		7 & Machinery and transport equipment\\
		8 & Miscellaneous manufactured articles\\
		9 & Commodities and transactions not classified elsewhere in the SITC\\
		\hline
	\end{tabular}
	\caption{\label{tab1}List of product groups from the Standard International Trade Classification (SITC) \cite{comtrade}.}
\end{table}

The first applications of Google matrix methods to the UN Comtrade based WTN
have been presented in \cite{benedictis,wtn1}. 
While only the PageRank algorithm was used in \cite{benedictis},
it was shown in \cite{wtn1} that it is equally important to use the CheiRank algorithm
which describes the outgoing trade flows.
Indeed, while the PageRank  describes the ingoing trade flows, ie, the import flows, the CheiRank, explained with details in \cite{linux1,wikizzs},
describes the export flows.
For the WTN study, it is clear that the combination of both the PageRank 
and the CheiRank analysis is of primary importance as it extends the traditional Import-Export analysis by taking account of the multiplicity of cascades of trade exchanges.  The further development of Google matrix 
applications to the WTN allowed to perform the analysis of the multiproduct WTN \cite{wtn2}
by considering the contributions of the products proportionally to their trade volumes and treating countries on equal grounds which 
corresponds to the fundamental principle of the UN.
These methods were also successfully applied to the networks of economics activities provided by the WTO \cite{kandiah15}.
In addition, the reduced Google matrix 
(REGOMAX) algorithm has been applied to the multiproduct WTN \cite{wtn3} and to the multisectorial trade network \cite{coquide20}.
This algorithm, developed and numerically tested in \cite{greduced,politwiki,diseases}, 
allows, for a subset of interest constituted by a moderate number $N_r$ of network nodes, to find the effective interactions between these  $N_r$ nodes taking into account all the 
indirect pathways via the global 
network of much larger size $N \gg N_r$ (for the here studied WTN case $N=N_c N_p = 194 \times 10 =1940$).
Using the above cited algorithms, we analyze here the impact of the COVID-19 on the WTN for the years 2018-2020.

It should be noted that there is a growing interest to the analysis of the COVID-19 influence on the WTN
in the last couple of years (see e.g. \cite{vidya,kiyota,zang,antoni}). However, these works were
concentrated on the analysis of the PageRank (analogous to the ImportRank) while the CheiRank (analogous to the ExportRank)
was not considered even if it is equally important. 
Also, the properties of the multiproduct trade were 
not analyzed. Thus, our global Google matrix analysis, including REGOMAX algorithm, 
allows to obtain a much deeper characterization of the COVID-19 impact on the WTN.

We note that the methods of statistical physics and network analysis in econophysics attract a growing interest
(see e.g. \cite{mantegna99,munnix14,bardoscia17,serrano07,deguchi14}) and we hope that our studies will bring new elements
to such investigations.
The WTN describes monetary flows of products between countries
that can be considered as information flows
related to the entropy which relates our studies to the fundamental aspects of the Markov chains and its statistical description.

\section{Data sets and Google matrix algorithms} 

We use the UN Comtrade data \cite{comtrade} 
for the years 2018, 2019 and 2020 (collected in September 2021) to 
construct the trade flows of the multiproduct WTN following \cite{wtn2,wtn3}
which give a detailed description of the applied methods. For each year, a money matrix, 
$M^{p}_{cc^{\prime}}$, gives the export flow of product $p$
from country $c^{\prime}$ to country $c$ 
(transactions are expressed in USD of the current year).
The money matrix element, $M_{cc'}^p$, is taken as the maximum between the reported volume of product $p$ imported by the country $c$ from the country $c'$ and the reported volume of product $p$ exported from the country $c'$ to the country $c$. Although these volumes should be of the same order, this procedure allows to reduce the effect of the absence or the delay of reporting by certain countries.
The data set concerns $N_{c} = 194$ countries and territories 
and $N_{p} = 10$ principal groups of products reported in Table~\ref{tab1}.
Thus, the total Google matrix $G$ size is  $N=N_c N_p= 1940$ corresponding to the number of couples (product, country).

The Google matrix elements $G_{ij}$ of direct trade flows
are constructed in a  way described with details in \cite{wtn2,wtn3}: 
the monetary values of products traded from a node $j$ to the other nodes $i$, renormalized in such a way that their sum is equal to one, constitute the elements of the matrix $S$ which represents
Markov trade transitions. Each column associated to a dangling node, ie, a node from which no trade flows emanates, is replaced by a column with all elements equal to $1/N$. A personalization vector $v$ encodes the relative weight of each product in the global trade volume and all the countries are treated on equal grounds 
following the main UN principle.
The Google matrix elements are $G_{ij}=\alpha S_{ij} + (1-\alpha) v_i$
where $v_i$ is the personalization vector component of node $i$ ($v_i>0, \,\forall i$ and $\sum_{i=1}^N v_i=1$).
Here, we use the damping factor value $\alpha=0.5$. 
Similarly, we also construct the Google matrix $G^*$ from the money matrix with inverted trade flows.

The stationary probability distribution on the WTN
is given by the PageRank vector
$P$ defined by the eigenproblem 
$G P=P$ \cite{brin,meyer,rmp2015}. 
For the WTN network with inverted trade flows, described by $G^*$, 
the stationary probability distribution is given by the CheiRank vector $P^*$, defined such as $G^* P^* = P^*$. 
The Page\-Rank node index $K$ and the Chei\-Rank node index $K^*$ are obtained by sorting in descending order the components of the PageRank vector $P$ and 
of the CheiRank vector $P^*$, respectively.
For the WTN, each node corresponds to a given couple $\left(p,c\right)$ of country $c$ and product $p$. Hence, the PageRank (CheiRank) component $P_{pc}$ ($P^*_{pc}$) measures the ability of a country $c$ to import (export) a product $p$. 
Following \cite{wtn2,wtn3}, the sums over all the product types $p$ give the PageRank probability $P_c =\sum_p P_{pc}$ and the CheiRank 
probability
$P^*_c =\sum_p P^*_{pc}$ of a given country $c$. Similarly, the sums over the countries provide the product $p$ PageRank and CheiRank probabilities, ie, $P_p=\sum_cP_{pc}$ and $P^*_p=\sum_cP^*_{pc}$. Sorting these probabilities in descending order allows us to define the related indexes
$K_c$,
$K^*_c$,
$K_p$,
and
$K^*_p$.
From export and import volumes, it is possible to define, for a couple $\left(p,c\right)$, an ImportRank probability $\hat{P}_{pc}=V^{-1}\sum_{c'}M_{cc'}^p$
and an ExportRank probability
$\hat{P}^*_{pc}=V^{-1}\sum_{c}M_{cc'}^p$, where $V=\sum_{p,c,c'}M^p_{cc'}$ is the total volume of the WTN. Analogously to the PageRank and CheiRank probabilities, it is possible to define the ImportRank and the ExportRank probabilities $\hat{P}_c$ and
$\hat{P}^*_c$ for the country $c$, and
$\hat{P}_p$,
and $\hat{P}^*_p$ for the product $p$. Also, we define $\hat{K}$ and $\hat{K}^*$ as the ImportRank and the ExportRank node index, and the other indexes
$\hat{K}_c$,
$\hat{K}^*_c$,
$\hat{K}_p$,
and
$\hat{K}^*_p$.
Qualitatively, the
PageRank probabilities are proportional to the volumes of ingoing
trade flow and the CheiRank probabilities to the volume of outgoing flow.
Thus, approximately, products and countries subject to high importations (exportations) have a high PageRank (CheiRank) probability.
Let us note, that contrarily to the usual import-export description, ie, the one obtained with the ImportRank and the ExportRank, which takes only account of bilateral direct trade exchanges, 
the PageRank and CheiRank description
takes into account the multiplicity of all the cascades of transactions encoded in the WTN.

Following \cite{wtn2,wtn3},
we determine the PageRank-CheiRank trade balance of a given country $c$ as
$B_c = (P^*_c - P_c)/(P^*_c + P_c)$.
In a similar way, the ImportRank-ExportRank trade balance is 
$\hat{B}_c=  ({\hat{P}^*}_c - \hat{P}_c)/({\hat{P}^*}_c + \hat{P}_c)$.
The sensitivity of the
trade balance $B_c$ to the price of a given product $p$ is obtained by multiplying the corresponding money matrix elements $\left\{M^p_{c'c''}\right\}_{\forall c',c''}$, by a factor $(1+\delta)$, where $\delta$ is an infinitesimal number. The PageRank-CheiRank trade balance sensitivity is then defined as $dB_c/d\delta$. 
It is also useful to determine the PageRank-CheiRank product $p$ trade balance defined as
$B_p = (P^*_p - P_p)/(P^*_p + P_p)$, which is, \textit{a priori}, non zero, contrarily to the ImportRank-ExportRank product $p$ trade balance
$\hat{B}_p = (\hat{P}^*_p - \hat{P}_p)/(\hat{P}^*_p + \hat{P}_p)$ 
which is equal to zero due to the  existing symmetry of product $p$
transactions $\hat{P}_p=\hat{P}^*_p$.
Finally, the product $p$ trade balances of a given country $c$ are defined as $B_{pc} = (P^*_{pc} - P_{pc})/(P^*_p + P_p)$ and
$\hat{B}_{pc}=  (\hat{P}^*_{pc} - \hat{P}_{pc})/({\hat{P}^*}_p + \hat{P}_p)$.

We also use the REGOMAX algorithm \cite{greduced,politwiki,wtn3}
to obtain the reduced Google matrix $\GR$ associated to a selected subset of $N_r$ WTN nodes (with $N_r \ll N$).
This algorithm allows to incorporate all the direct and
indirect transaction pathways connecting any couples of nodes among the $N_r$ nodes of interest.
The reduced Google matrix $\GR$, obtained from the full WTN Google matrix $G$, is used
to construct a reduced network summarizing the strongest trade interactions
between a selection of nodes representing countries and products.

More detailed descriptions of the Google matrix based algorithms and the related numerical methods
can be found in \cite{wtn1,wtn2,wtn3,greduced,politwiki}.

Let us note that the above depicted algorithms have been used in \cite{wtn4,wtn5}
to analyze the multiproduct WTN for years before 2019, and hence, the
COVID-19 effects on the WTN have not been captured there. 

\section{Results}

\begin{figure}[!t]
	\centering
	\includegraphics[width=\textwidth]{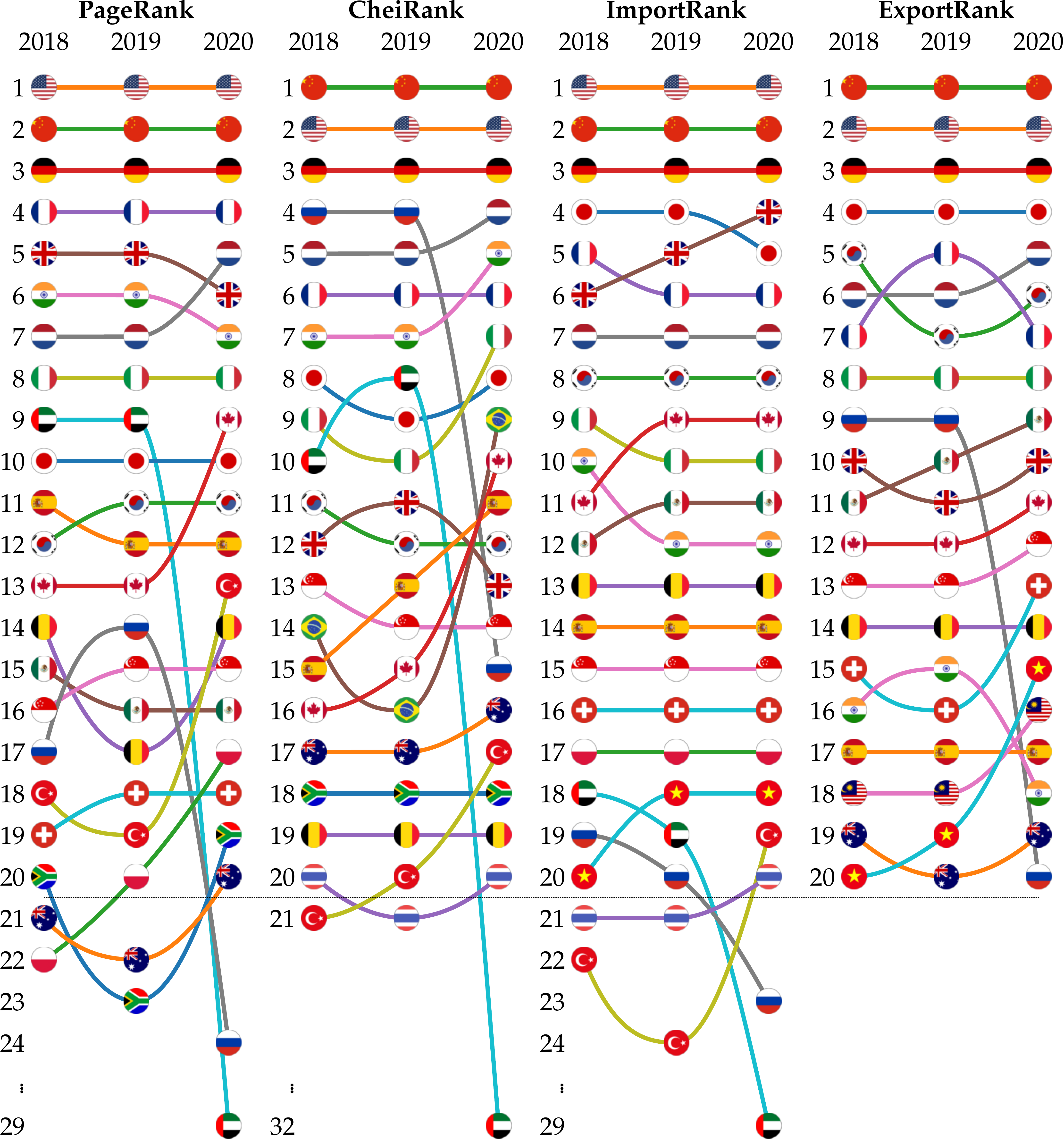}
	\caption{\label{tab2}Top20 of countries ranked according to the PageRank, CheiRank, 
		ImportRank, and ExportRank algorithms for the years 2018, 2019 and 2020. The countries are represented by their national flags.}
\end{figure}

\subsection{World countries on the PageRank-CheiRank plane}
The top20 countries according to the PageRank, CheiRank, ImportRank and ExportRank
are given in Fig.~\ref{tab2} for the years 2018, 2019 and 2020.
For each ranking, the top3 countries are unchanged for the three consecutive years. The top3 countries are either  USA, China and Germany for import related rankings, ie, ImportRank and PageRank, or China, USA and Germany for export related rankings, ie, ExportRank and CheiRank.
Globally, along this time period, the Import\-Rank and the Export\-Rank are quite stable for the top20 main international trade players suggesting that their relative total trade capacities were not significantly affected by the COVID-19.
Indeed, from 2018 to 2020, we observe 10 countries for which the rank does not change while the rank of the others change by at most $\pm2$ positions. However, there are the following exceptions: Russia which significantly worsen both its ImportRank ($\hat{K}=19\rightarrow23$) and ExportRank ($\hat{K}^*=9\rightarrow20$), the United Arab Emirates (UAE) which worsen its ImportRank ($\hat{K}=18\rightarrow29$, UAE does not even appear in the top20 exporters) and Vietnam which improves its ExportRank ($\hat{K}^*=20\rightarrow15$).
The relative loss of export volume for Russia and UAE can be attributed to the significant global reduction of the world production and, also, of the air traffic, induced by the COVID-19 lockdown. It reduced automatically the world request for  mineral fuels (petrol, gas\dots) which dominate trade flows of Russia and UAE. On the opposite, the Vietnamese economy was one of the most resilient during the COVID-19 crisis \cite{bbcvietnam}.

Contrarily to the import-export description, based only on the relative amount of total export and import volumes, the PageRank and CheiRank description takes account of the structure of the WTN and consequently allows to rank countries according to their relative efficiency to import and export commodities. Roughly speaking, countries importing from (exporting to) a large variety of countries will obtain a better PageRank (CheiRank) than ImportRank (ExportRank).
Eg, India, in 2018, which has the 16th export volume (10th import volume), occupies the $K^*=7$ ($K=6$) position in the CheiRank (PageRank) suggesting a more important role in the WTN structure as it would be inferred by focusing only on the import-export point of view. We can observe for 2018 similar cases for UAE ($\hat{K}=18\rightarrow K=9$, $\hat{K}^*=21\rightarrow K^*=10$) and for Russia ($\hat{K}^*=9\rightarrow K^*=4$).
Taking the example of Russia, the diversity of its trade relations for mineral fuels, involving both European and Asian countries, is much better captured by the PageRank-CheiRank picture which takes account of the complexity of the WTN and the multiplicity of cascades of commercial links.
Conversely, despite that Japan's import and export volumes are the 4th most important in 2018, Japan's position drops to $K=10$ in PageRank and $K^*=8$ in CheiRank suggesting a weaker centrality than expected of Japan in the WTN certainly due to a lack of diversity of commercial partners.

The PageRank-CheiRank picture allows to detect a rewiring of the WTN due to the COVID-19 pandemic. Indeed, from
Fig.~\ref{tab2} (two first columns), we observe more abrupt variations from 2019 to 2020
(up to +6 and -20 positions for the PageRank of Turkey and UAE, respectively;
up to +7 and -22 positions for the CheiRank of Brazil and UAE, respectively)
than from 2018 to 2019
(up to +3 and -3 positions for the PageRank of Russia and Belgium, respectively;
up to +2 and -2 positions for the CheiRank of, eg, Spain and Brazil, respectively).
Mimicking the loss of import and export volumes experienced by Russia and UAE from 2018 to 2020, the Russia's CheiRank (PageRank) drops by 11 (7) positions and the UAE's CheiRank (PageRank) drops by 22 (20) positions. Excepting these two cases explained by extreme variations of trade volumes, we observe anyway for some other countries quite large variations from 2018 to 2020. Eg, Brazil gains 5 positions in CheiRank climbing up to $K^*=9$ whereas its volume of export stays stable around the 24th position ($\hat{K}^*=24$ in 2018 and 2019, $\hat{K}^*=23$). A similar gain is observed for Canada ($K^*=16\rightarrow10$) but with a higher export volume ($\hat{K}^*=12$ in 2018 and 2019, $\hat{K}^*=11$).
India also gain +2 CheiRank positions which can be attributed to its broad trade connections with Asia, Europe, America and an enhanced export of variety of products, e.g. pharmaceutics during COVID-19.
Among European countries, we note that the Netherlands, Italy, Spain improved their CheiRank positions from 2018 to 2020.

\begin{figure}[th]
		\centering
		\includegraphics[width=\textwidth]{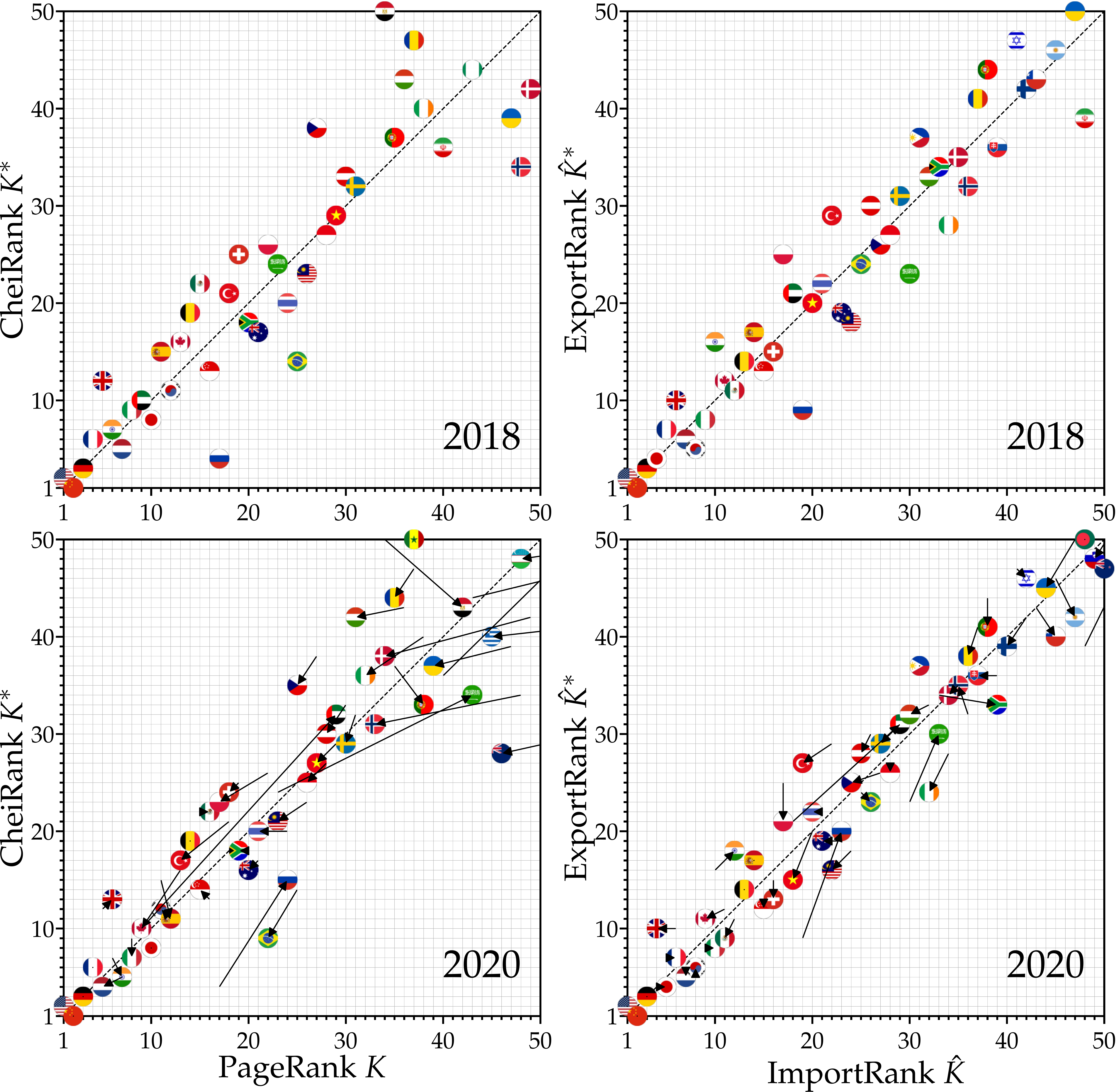}
	\caption{\label{fig1}Distributions of countries on 
		the PageRank $K$ - CheiRank $K^*$ plane (left panels) and 
		on the ImportRank $\hat K$ - ExportRank $\hat{K}^*$ (right panels). 
		The distribution are shown for the years 2018 (top panels) and 2020 (bottom panels). 
		On the bottom panels, each arrow indicates the change between the two years: 
		the arrow tail [head] is located on the point $\left(K,K^*\right)$ for 2018 [2020]. 
		The countries are represented by their corresponding national flags.
	}
\end{figure}

A more synthetic view of the COVID-19 effects on the rankings can be observed on the PageRank-CheiRank and ImportRank-ExportRank planes displayed in Fig.~\ref{fig1}.
Globally, the countries are located in the vicinity of the diagonal $K=K^*$ or $\hat{K}=\hat{K}^*$. Indeed, each country tries to keep, at the best, a balance between its import and export volumes. Countries located below the diagonal, ie, those with $K^*<K$ or $\hat{K}^*<\hat{K}$, are better exporters than importers. Conversely, countries above the diagonal, ie, those with $K^*>K$ or $\hat{K}^*>\hat{K}$, have better import than export capabilities.
Besides providing a summary of the previously discussed results obtained from the top20s displayed in Fig.~\ref{tab2}, the $(K,K^*)$-plane or the $(\hat{K},\hat{K}^*)$-plane views with $K,K^*,\hat{K},\hat{K}^*\in\left\{1,\dots,50\right\}$ allow directly to observe on a broader range the effects of the COVID-19. From Fig.~\ref{fig1}, the ranking changes observed from 2018 to 2020 are highlighted by arrows; the tails (tips) of which show the position of countries in 2018 (2020). The Fig.~\ref{fig1}, bottom right panel, shows that the import and export volumes does not vary much during the COVID-19 crisis with the exceptions of few cases such as, eg, Russia and Saudi Arabia which strongly worsen their relative import-export trade balance, or Vietnam and Ireland which improve it. The changes are larger in the PageRank-CheiRank plane (Fig.~\ref{fig1}, bottom left panel), highlighting a large scale rewiring of the WTN as a response to the COVID-19.
In addition, Fig.~\ref{fig1bis} in the Appendix, clearly shows that, during a pre-COVID-19 period, here from 2018 to 2019, by contrast with the period 2018-2020 (Fig.~\ref{fig1}), there are no significant variations of country positions on the PageRank-CheiRank plane and in the ImportRank-ExportRank plane.

\begin{figure}[!ht]
	\begin{center}
		\includegraphics[width=0.75\textwidth]{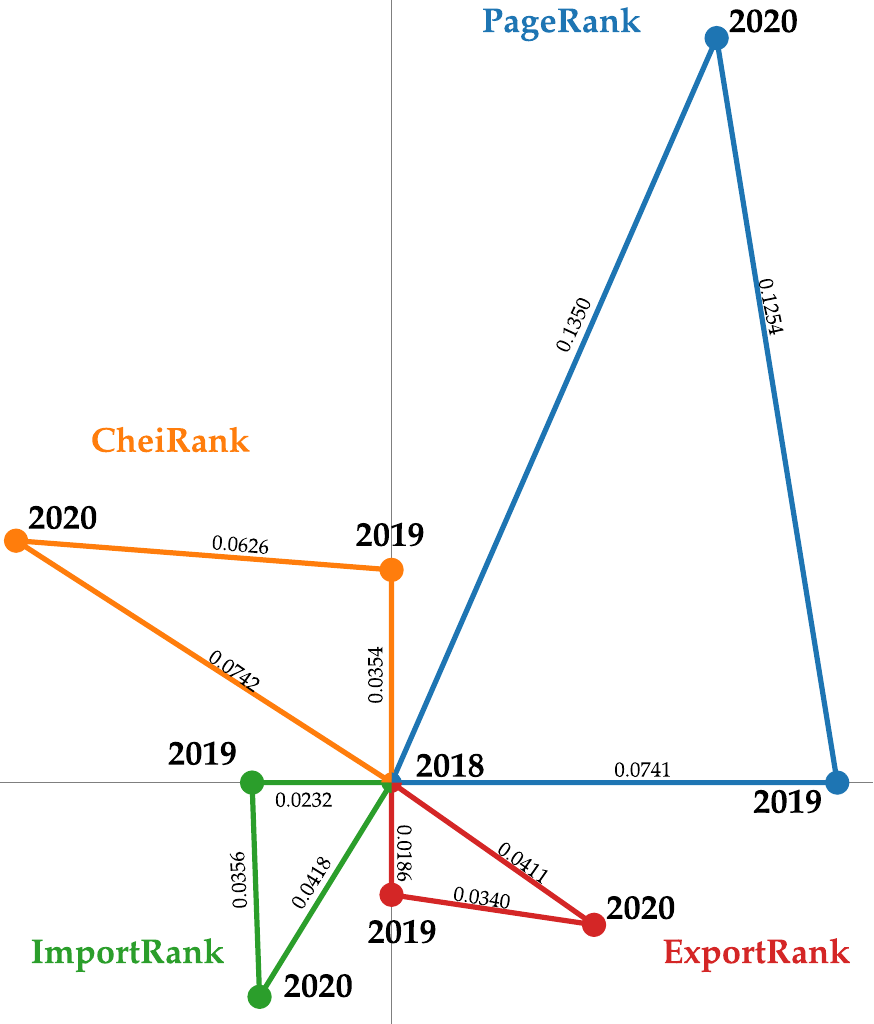}
	\end{center}
	\caption{\label{distance}Kendall $\tau$ distance between 2018, 2019, and 2020 ranking lists of the 194 world countries.
		The blue, orange, green and red points correspond to the PageRank, CheiRank, ImportRank and ExportRank country lists, respectively. The Kendall $\tau$ distance between ranking lists are mentioned along the segments.
		For the four rankings, the points corresponding to the 2018 lists are superimposed at the center of the picture.}
\end{figure}

The Kendall $\tau$ distances (see definition in Appendix~\ref{app:distance}) between the ranking lists for the different considered years are shown in Fig.~\ref{distance}. We clearly see the COVID-19 influence, since
for all the rankings, the 2018-2020 distance and the 2019-2020 distance are on average similar but they are two times greater than the pre-COVID-19 2018-2019 distance.
For import (export) metrics, we observe that the distances computed between PageRank (CheiRank) lists are, on average, 3.3 (1.8) greater than those computed between ImportRank (ExportRank) lists. This observation supports again the fact that COVID-19 influence is better probed by the Google matrix based rankings. These results are corroborated by the fact that the number of links decreases by about 20\% from 2018 to 2020 whereas the total volume exchanged decreases by about 14\%. From 2018 to 2019, these two quantities evolved similarly and lost about 3\%.
In addition, we note that although the ExportRank and the ImportRank triangles have similar size, the PageRank triangle is about 2 times greater than the CheiRank triangle. This suggests that the rewiring induced by the COVID-19 affects in a stronger manner the set of importing countries than exporting countries.

\begin{figure}[!ht]
		\centering
		\includegraphics[width=\textwidth]{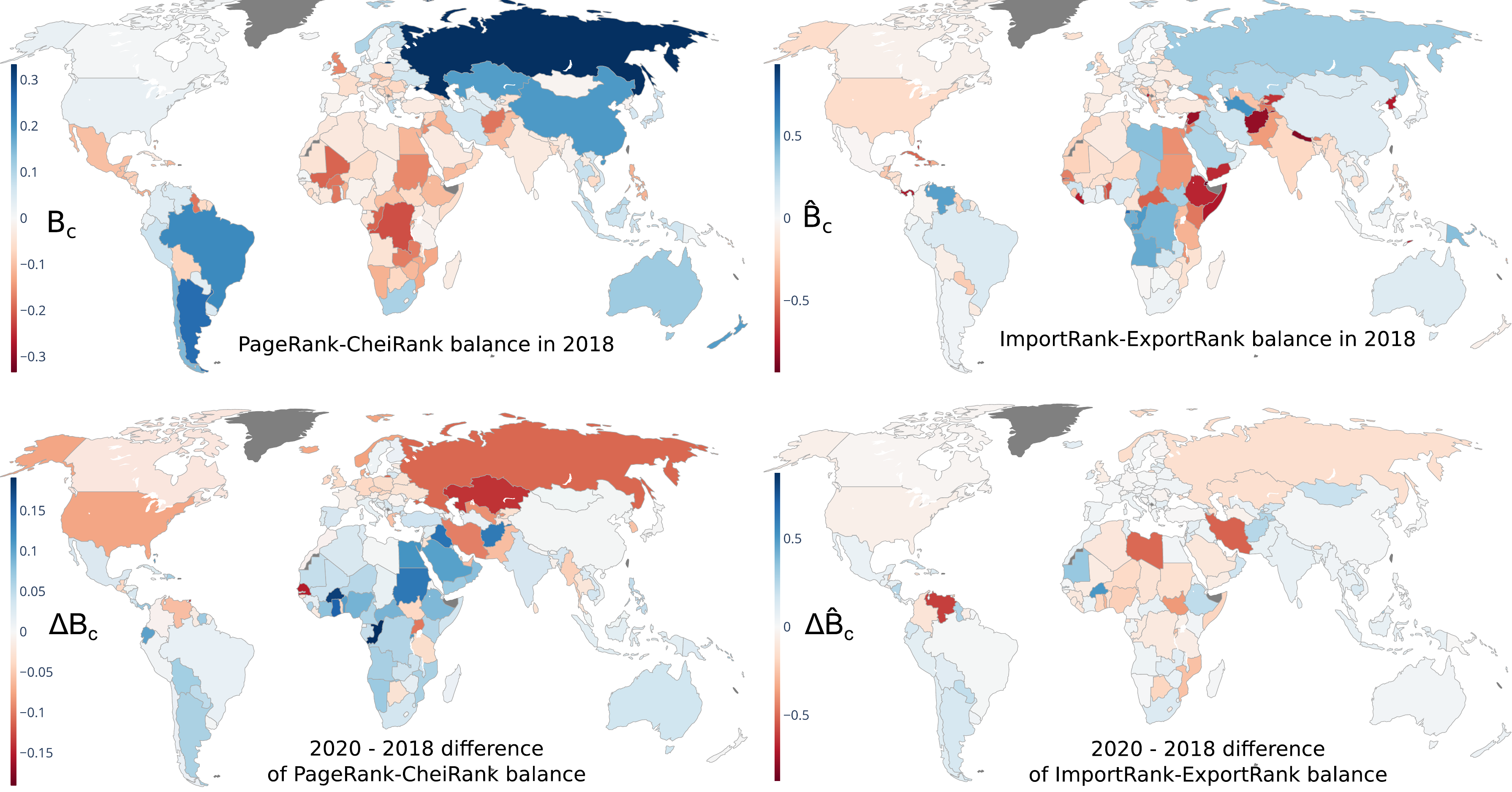}
	\caption{\label{fig2}Geographical distributions of the PageRank-CheiRank trade 
		balance $B_c$ (left panels) and of the ImportRank-ExportRank 
		trade balance $\hat B_c$ (right panels). The trade balances for the year 2018 
		are presented on the first row. The second row displays the difference of 
		trade balance between the years 2018 and 2020, ie, 
		$\Delta B_c=B_c^{(2020)}-B_c^{(2018)}$ (bottom left panel) and
		$\Delta  \hat B_c=\hat B_c^{(2020)}-\hat B_c^{(2018)}$ (bottom right panels).
	}
\end{figure}

\subsection{Trade balances and the COVID-19 influence} The geographical distributions of the PageRank-CheiRank trade balance $B_c$ and of the ImportRank-ExportRank trade balance $\hat{B}_c$ 
are shown in Fig.~\ref{fig2} for the year 2018. The variation from 2018 to 2020 are also displayed. In 2018, the strongest positive PageRank-CheiRank trade balances $B_c$ are obtained (by descending order)
for
Russia ($B_{\rm RU}\simeq0.33$),
Argentina, 
Brazil,
China,
Kazakhstan,
New Zealand,
Chile,
Australia,
Fiji,
South Africa,
and Norway ($B_{\rm NO}\simeq0.1$). These countries, several of which belong to the BRICS group, are more efficient exporters than importers of commodities, and consequently have a more robust international trade economy than countries with smaller or even negative PageRank-CheiRank trade balance.
The strongest negative PageRank-CheiRank trade balance are obtained for
the Democratic Republic of the Congo ($B_{\rm CD}\simeq-0.22$),
the Republic of the Congo ($B_{\rm CG}\simeq-0.2$)
and
Mali ($B_{\rm ML}\simeq-0.19$).
Contrarily to the ImportRank-ExportRank trade balance geographical distribution (Fig.~\ref{fig2}, top right panel) which appears inhomogeneous, with strong variations of balance between neighboring countries, the PageRank-CheiRank trade balance geographical distribution (Fig.~\ref{fig2}, top left panel) appears homogenous at the scale of a continent. For the latter distribution, we observe that negative balances $B_c<0$ extend over the African continent (excepted South Africa), over the region comprising Mexico, the Caribbean, and the Central America, from the Middle East (excepted Iran) to South Asia, Central Europe (excepted Germany).
We note also that the PageRank-CheiRank trade balance of the UK is negative, $B_{\rm UK}\simeq-0.16$.
The positive balance $B_c>0$ regions are mainly the Russia-Kazakhstan-China region, the Oceania, the South America (excepted Bolivia and the Guianas), and with lesser importance North America and the North of Europe. A partial explanation to the homogeneity of the positive and negative PageRank-CheiRank trade balance regions could be that an economically virtuous (non virtuous) country has in general developed trade relations with its neighbors which take advantage (disadvantage) of the good (bad) economic health of the former. This is a partial explanation in the current globalized world since long distance trade exchanges are, at the least, as important as regional trade exchanges.

The changes $\Delta B_c$ and  
$\Delta \hat{B_c}$ of balances from 2018 to 2020 are shown in the bottom panels
of Fig.~\ref{fig2} (left and right). 
We clearly see from the PageRank-CheiRank trade balance variations (Fig.~\ref{fig2}, bottom left) that countries which the economic health is the most affected by the COVID-19 are, in general, the developed countries \cite{wikideveloped} (with the major exception of Australia, New Zealand, Japan, Portugal, Spain, Italy, Baltic states, Sweden, and Finland) and many economies strongly dependent on oil and gas exports, such as those of Russia, Iran, UAE, Venezuela, and Trinidad and Tobago which, the latter, have the most affected balance. By contrast, many developing and undeveloped (mostly African, South Asian, and Sud American countries) increase their PageRank-CheiRank trade balance $B_c$. This result is due to the slow down of the global economy  driven by the developed countries. Indeed, the variety of the import trades from developed countries towards developing and undeveloped countries significantly decreased from 2018 to 2020 automatically enhancing their PageRank-CheiRank trade balance.
The most negative ImportRank-ExportRank trade balance variation $\Delta\hat{B}_c$ is found (by ascending order) for Venezuela ($\hat{B}_{\rm VE}\simeq-0.6$), Iran, Lybia, and South Sudan ($\hat{B}_{\rm SS}\simeq-0.38$) indicating crudely the bad economic health of these countries.

We argue that the PageRank-CheiRank approach provides a more adequate description of trade balance taking into account the multiplicity of the trade links in contrast to the gobal volume exchanged description given 
by the import-export standard approach conventionally used 
in the international trade.

\begin{figure}[!ht]
	\centering		\includegraphics[width=\textwidth]{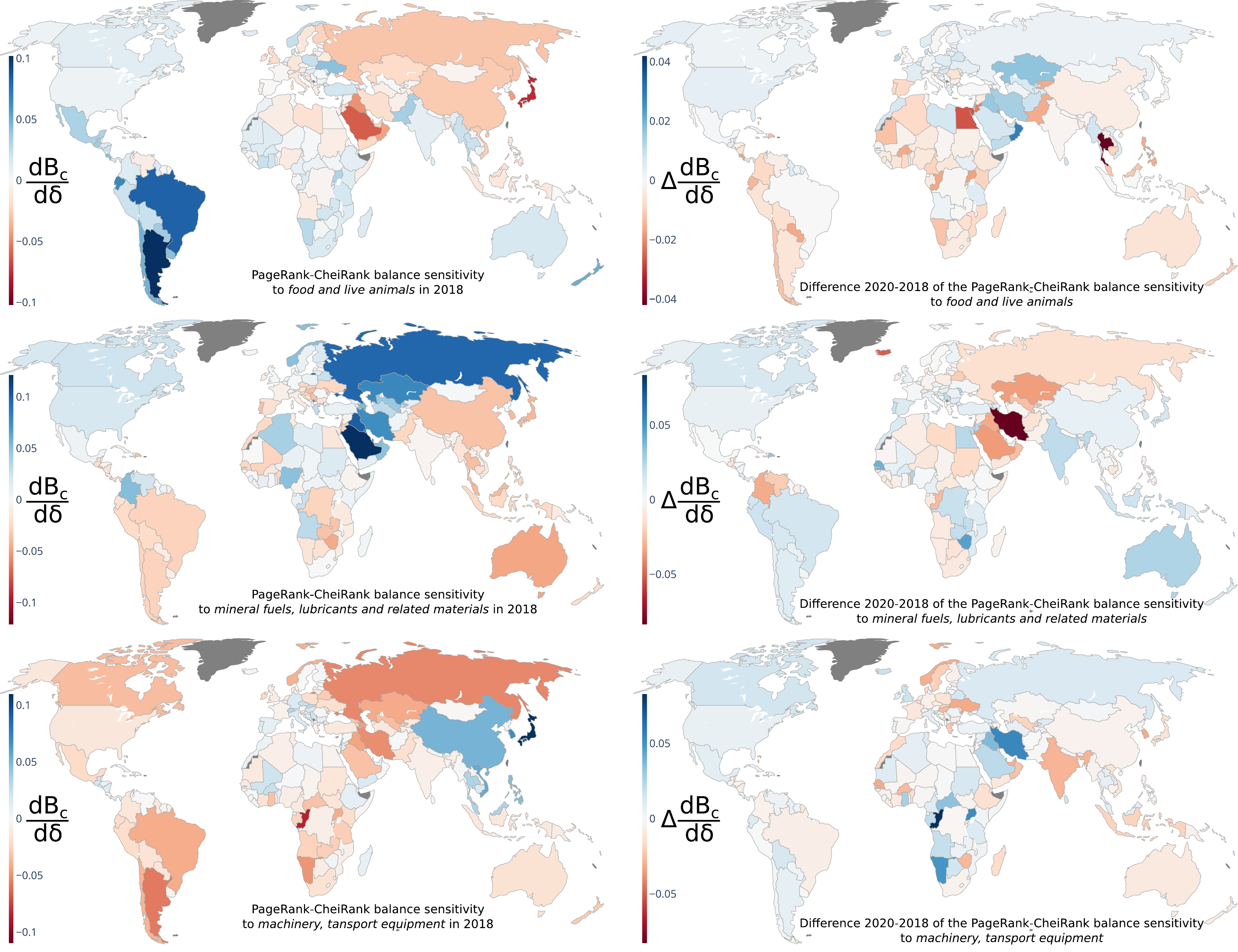}
	\caption{\label{fig3}Geographical distributions of the PageRank-CheiRank trade 
		balance sensitivity $dB_c/d\delta$ to \textit{food and live animals} ($p=0$, top row),
		\textit{mineral fuels, lubricants and related materials} ($p=3$, middle row), and
		\textit{machinery, transport equipment} ($p=7$, bottom row) products.
		The left column presents the PageRank-CheiRank trade balance sensitivities for the year 2018.
		The right column presents the difference between the years 2020 and 2018, ie,
		$\Delta dB_c/d\delta=\left(dB_c/d\delta\right)^{(2020)}-\left(dB_c/d\delta\right)^{(2018)}$.
	}
\end{figure}

\subsection{Trade balance sensitivity to product prices} The left column of Fig.~\ref{fig3}
shows the sensitivity $dB_c/d\delta$ of the country trade balance $B_c$ to a price increase
of a group of products $p=0$, $3$, and $7$ from Table~\ref{tab1}; the right column of Fig.~\ref{fig3}
gives the variation of sensitivity  $\Delta dB_c/d\delta = (dB_c/d\delta)^{(2020)} -  (dB_c/d\delta)^{(2018)}$
between years 2018 and 2020.

For {\it food and live animals} ($p=0$ in Tab.~\ref{tab1}), the largest positive balance sensitivities $dB_c/d\delta$
are found for South and Central America, especially, (by ascending order, for Argentina with $dB_{\rm AR}/d\delta\simeq0.1$, Brazil, Ecuador, Chile, and Uruguay with $dB_{\rm UY}/d\delta\simeq0.04$), New Zealand ($dB_{\rm NZ}/d\delta\simeq0.05$), and Ukraine ($dB_{\rm UA}/d\delta\simeq0.04$),
whereas the largest negative 
balance sensitivities $dB_c/d\delta$
are found for Japan ($dB_{\rm JP}/d\delta\simeq-0.08$), Arabian Peninsula countries (such as Saudi Arabia with $dB_{\rm SA}/d\delta\simeq-0.06$, Kuwait, Iraq, and Oman with $dB_{\rm OM}/d\delta\simeq-0.04$), and Russia and China with, $dB_{\rm RU}/d\delta\simeq dB_{\rm CN}/d\delta\simeq-0.03$. The former (latter) set of country would have benefited from (been disadvantaged by) a price or a volume exchanged increase of \textit{food and live animals}.
We attribute the variation of the sensitivity  $\Delta dB_c/d\delta$ to COVID-19
which induced a positive increase of the balance trade sensitivity for Oman, Iraq, Iran, and Kazakhstan, 
and a negative decrease for Thailand, Egypt, and Jordan.

For {\it mineral fuels, lubricants and related materials} ($p=3$ in Tab.~\ref{tab1}), 
the largest positive balance sensitivities are found (by descending order) for
Saudi Arabia ($dB_{\rm SA}/d\delta\simeq0.12$),
Iraq, Russia, Kuwait, Qatar, Kazakhstan, Iran, Azerbaijan, and UAE ($dB_{\rm AE}/d\delta\simeq0.53$)
which are world leading petroleum and gas producers. By contrast, the negative sensitivities are at the least equal to $-0.047$ for Australia and are distributed mostly over all the countries which are not major oil and petroleum exporters.
The largest positive variation $\Delta dB_c/d\delta$
is obtained, among the large country, for Australia, India, Egypt that we attribute to 
a reduction of petroleum import due to the COVID-19 production reduction.
The largest negative variation $\Delta dB_c/d\delta$ is for
Iran, Kazakhstan, and Saudi Arabia which suffered from petroleum demand reduction.

For {\it machinery and transport equipment} ($p=7$ in Tab.~\ref{tab1}), the largest positive balance sensitivities are obtained (by descending order) 
for Japan ($dB_{\rm JP}/d\delta\simeq0.11$), South Korea, Vietnam, and China ($dB_{\rm CN}/d\delta\simeq0.052$);  the  largest negative values are obtained (by ascending order) for Congo ($dB_{\rm CG}/d\delta\simeq-0.085$), Argentina, Cyprus, Russia, and Iran ($dB_{\rm IR}/d\delta\simeq-0.051$).
We also observe a positive relative sensitivity to \textit{machinery and transport equipment} for European countries. Many of these countries in addition to India and South Korea have a negative variation of the sensitivity due to COVID-19.

For reader convenience, the geographical distribution of the trade balance sensitivity obtained from the ImportRank and ExportRank is also given in the Appendix (see
Fig.~\ref{figS1}). We do not discuss here this figure since as explained above this approach is less informative. 

\begin{figure}[!ht]
	\begin{center}
		\includegraphics[width=0.75\textwidth]{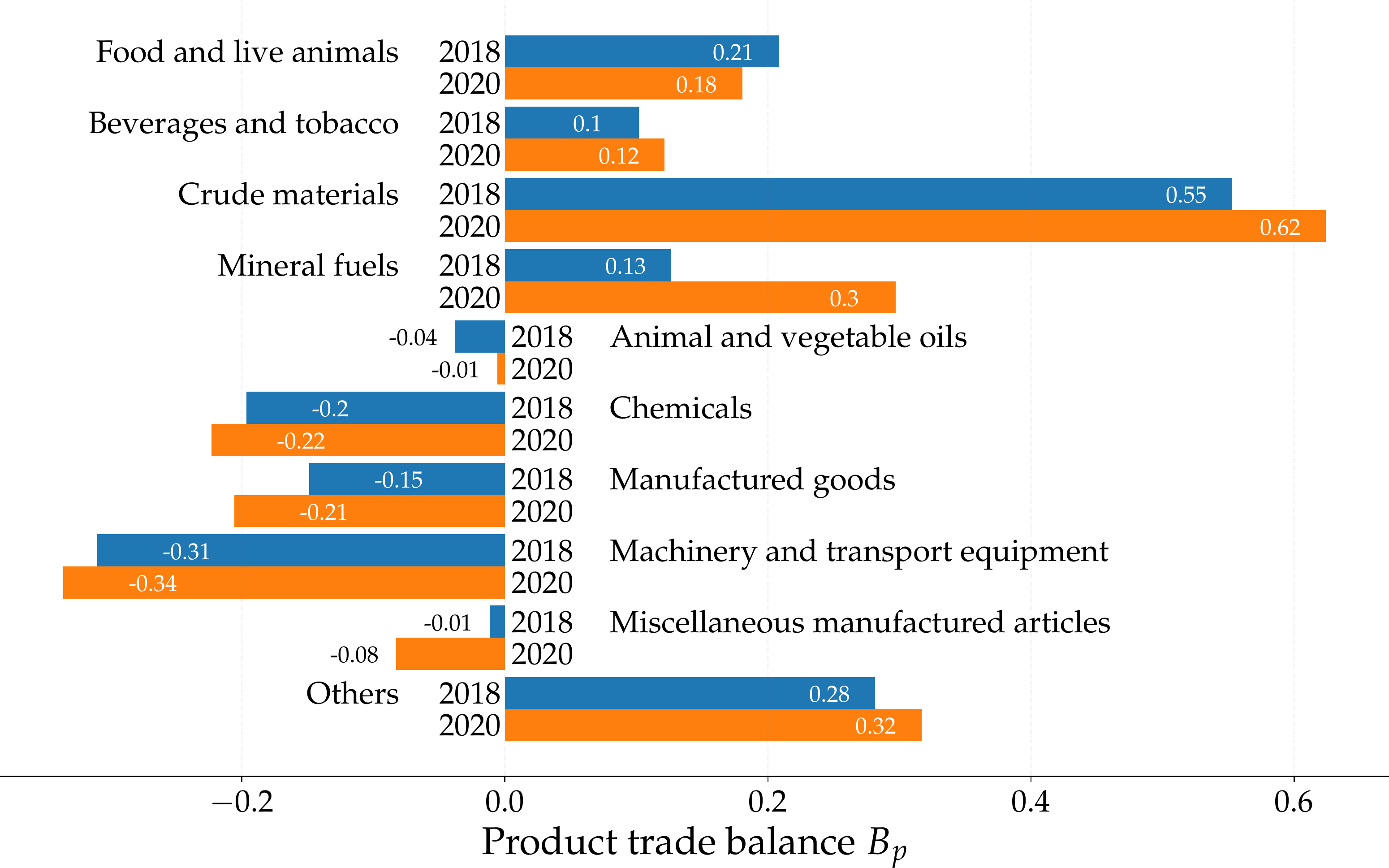}
	\end{center}
	\caption{\label{fig4}PageRank-CheiRank product trade balance $B_p$ 
		for years 2018 (blue bars) and 2020 (orange bars); the product balance $\hat{B}_p$ computed from ImportRank-ExportRank is zero.
	}
\end{figure}

\subsection{PageRank-CheiRank product trade balance} It is interesting to analyze the trade balance not only for countries but also
for products performing summation of $B_{pc}$ over countries. This peculiarity is not possible in the case of the ImportRank-ExportRank approach, since, by definition,
the ImportRank-ExportRank product trade balance is zero, $\hat{B}_p=\sum_c\hat{B}_{pc}=0$. Indeed, after summation over countries, the Export and Import probabilities, ie, the relative export and import volumes, are equal, ${\hat{P}^*}_{p} =\hat{P}_{p}$. However, within the PageRank-PageRank approach,
the probabilities $P^*_{p}$ and $P_{p}$ are different, and, thus, the product trade balance of some products, $B_p=\sum_cB_{pc}$, can be positive, and negative for some others. The trade balances for each group of products (Tab.~\ref{tab1}) is presented in Fig.~\ref{fig4}
for the years 2018 and 2020. The balance $B_p$ is positive for products $p=0, 1, 2, 3, 9$
and hence they are more export oriented, $P^*_p > P_p$. These products are more central in the inverted WTN than in the direct WTN. Otherwise stated, the export abilities of these products exceeds their import abilities.
In contrast, the products
$p=4, 5, 6, 7, 8$ have negative balances and are import oriented, $P_p > P^*_p$. These products import abilities supersede export abilities.
The PageRank-CheiRank approach allows to infer a kind of structural law of supply and demand.
The strongest variation of product balance from 2018 to 2020 takes place for 
{\it mineral fuels} ($p=3$) with an increase of $B_p$ by a factor $2.3$ from 2018 to 2020,
while the strongest negative decrease of $B_p$ is observed for {\it manufactured goods} ($p=6$)
by factor $1.4$ from 2018 to 2020. An even stronger negative decrease of $B_p$ takes place for
{\it miscellaneous manufactured articles} ($p=8$) with a factor $8$, however for this
product the balance $B_p$ is relatively small.
The effects of the COVID-19 are salient here. The products \textit{manufactured goods} and \textit{miscellaneous manufactured articles} lose export ability during 2020 whereas import ability were maintained. Conversely, export abilities were maintained for \textit{mineral fuels} but structural demand has declined in 2020.

\begin{figure}[!ht]
		\centering
		\includegraphics[width=0.75\textwidth]{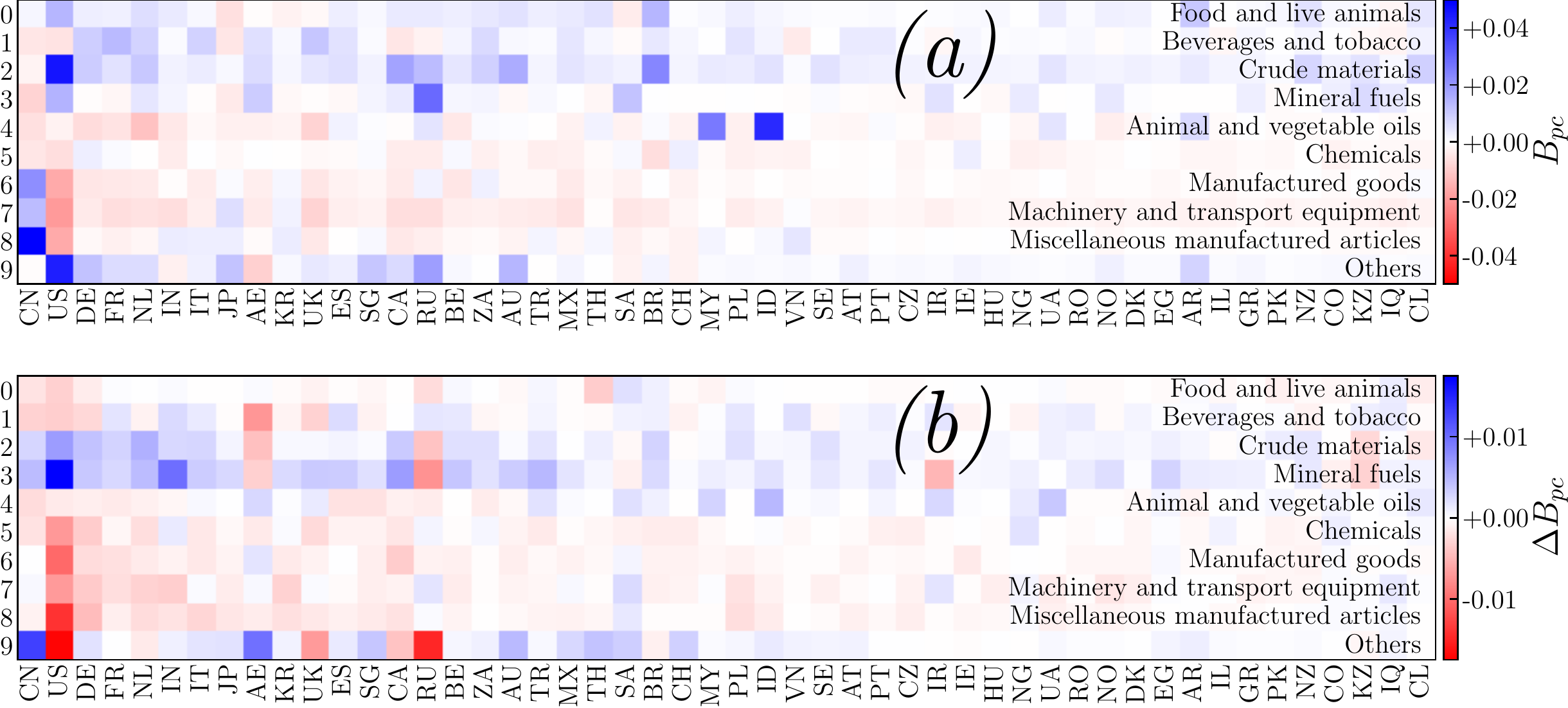}
	\caption{\label{fig5}PageRank-CheiRank (product,country) trade balance $B_{pc}$ [panel $(a)$]  
		for the year 2018.  Trade balance differences between years 2020 and 2018, 
		$\Delta B_{pc}=B_{pc}^{(2020)}-B_{pc}^{(2018)}$ are presented in the panel $(b)$. 
		The columns correspond to countries $c$, sorted according to 2DRank (see Tab.~\ref{tab3} for the first 20), and 
		rows correspond to products $p$ (see Tab.~\ref{tab1}).
		The meaning of the ISO 3166-1 alpha-2 country codes not present in Tab.~\ref{tab3} can be found, eg, at \cite{wikicc}.
	}
\end{figure}

The distribution of product trade balance by countries, $B_{pc}$, obtained
from the PageRank-CheiRank approach, is shown in Fig.~\ref{fig5} 
for 2018. The variations, $\Delta B_{pc}$, from 2018 to 2020 of the product trade balances per country are also shown
(the results obtained from the ImportRank-ExportRank approach are given in the Appendix, see Fig.~\ref{figS2}).
In 2018, China has positive trade balance $B_{pc}$ for products $p=6$, $7$ and $8$, with the most negative trade balance 
$B_{pc}$ for product $p=3$. The product trade balance variation from 2018 to 2020 is significantly positive for
product $p=9$ (which includes financial transactions) and slightly positive (negative) for $p=3$  ($p=2$ and $1$).
The COVID-19 related variation of USA product trade balance from 2018 to 2020
is significantly negative for products $p=4$, $5$, $7$, $8$, $9$ and positive for products $p=2$, $3$.
Among  significant positive values of $B_{pc}$ for the other countries in Fig.~\ref{fig5}a
we point: Indonesia and Malaysia for $p=4$,  Russia for $p=3$ and Brazil for $p=2$;
the negative product trade balances being less pronounced. 
Among the most significant  variations $\Delta B_{pc}$ in Fig.~\ref{fig5}(b),
for other countries, we point: India with a positive value for $p=3$, United Arab Emirates 
with positive $p=9$ and negative $p=1$; negative variation for Russia with $p=9$, $3$, $2$. For the reader interest, the same figure as Fig.~\ref{fig5} but obtained from ImportRank-ExportRank can be found in the Appendix (see Fig.~\ref{figS2}).

\begin{table}[!ht]\centering
	\begin{tabular}{|r|r|r|c|l|}
		\hline
		$K_2$ & $K$ & $K^*$ & CC & Country\\
		\hline
		\hline
		1 & 2 & 1 & CN & China\\
		2 & 1 & 2 & US & United States\\
		3 & 3 & 3 & DE & Germany\\
		4 & 4 & 6 & FR & France\\
		5 & 7 & 5 & NL & Netherlands\\
		6 & 6 & 7 & IN & India\\
		7 & 8 & 9 & IT & Italy\\
		8 & 10 & 8 & JP & Japan\\
		9 & 9 & 10 & AE & United Arab Emirates\\
		10 & 12 & 11 & KR & South Korea\\
		11 & 5 & 12 & UK & United Kingdom\\
		12 & 11 & 15 & ES & Spain\\
		13 & 16 & 13 & SG & Singapore\\
		14 & 13 & 16 & CA & Canada\\
		15 & 17 & 4 & RU & Russia\\
		16 & 14 & 19 & BE & Belgium\\
		17 & 20 & 18 & ZA & South Africa\\
		18 & 21 & 17 & AU & Australia\\
		19 & 18 & 21 & TR & Turkey\\
		20 & 15 & 22 & MX & Mexico\\
		\hline
	\end{tabular}
	\caption{\label{tab3}List of top 20 most efficient exporting 
		or importing countries according to the PageRank and the CheiRank algorithms.
		These 20 countries are located in the $\left[1,22\right]\times\left[1,22\right]$ 
		square of the $\left(K,K^*\right)$-plane (see Fig.~\ref{fig1}). 
		These 20 countries are sorted using the 2DRank indexes $K_2$ \cite{wikizzs}, 
		ie, by ascending value of $\max\left(K,K^*\right)$ and then, in case of \textit{ex-\ae quo}, 
		by ascending value of $K^*$. The column CC gives the ISO 3166-1 alpha-2 
		country codes (see eg \cite{wikicc}).}
\end{table}

\subsection{COVID-19 induced rewiring of the WTN} In Fig.~\ref{figS3}, 
we show the three reduced Google matrices $\GR$
for the subset of 20 countries with top PageRank and CheiRank indexes (see Tab.~\ref{tab3}) 
for 3 selected products $p=0, 3, 7$ in year 2018. 
Such a  matrix $\GR$ describes the effective trade flows of a given product between 
these 20 countries.  We also present the variation $\Delta \GR = \GR^{(2020)} - \GR^{(2018)}$ 
of the matrix elements between 2018 and  2020 which we attribute to COVID-19 impact. Indeed, the variations over a year in the pre-COVID-19 period are two to five less significant than the variations between 2018 and 2020 (see Fig.~\ref{figGR1819} in the Appendix).
The above described REGOMAX algorithm
takes into account all the indirect chains of trade transactions between these 20 countries
in the global WTN (1940 nodes, 449675 trade transactions over the considered period).

\begin{figure}[!ht]
		\centering
		\includegraphics[width=0.65\textwidth]{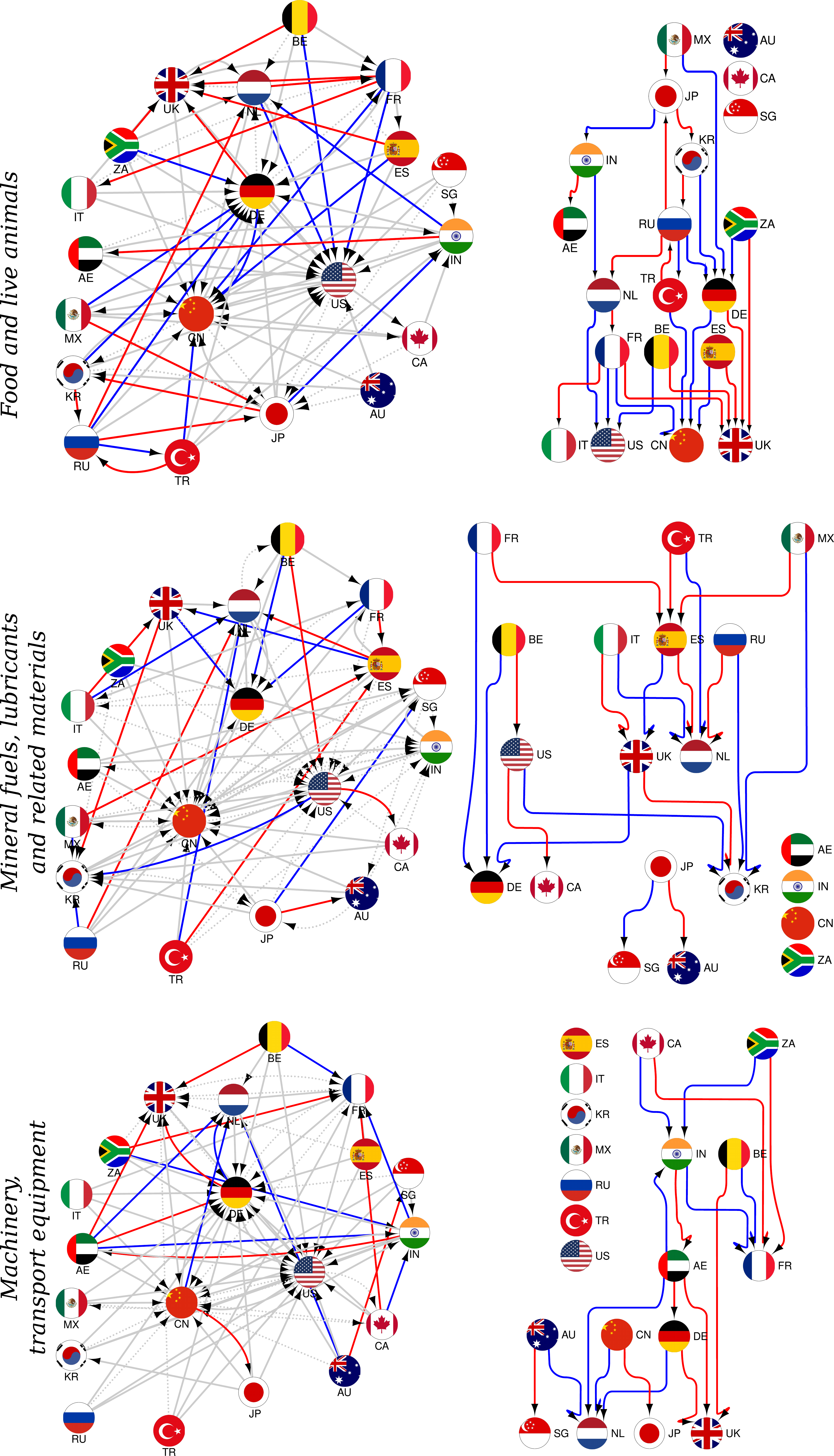}
	\caption{\label{fig6} Reduced networks for
		\textit{food and live animals} (top),
		\textit{mineral fuels, lubricants and related materials} (middle), and
		\textit{machinery, transport equipment} (bottom) trade exchanges.
		On the left panel, the reduced networks of the top 20 most efficient exporters 
		or importers according to the PageRank and the CheiRank algorithms (see Tab.~\ref{tab3}). For each panel, data from the years 2018 and 2020 are combined.
		For each year, and from each country, the four most important export transitions 
		encoded in the $G_{\rm R}$ are kept. The gray arrows correspond to important export transitions 
		existing in both years 2018 and 2020. The blue arrows correspond to important export transitions 
		appearing in 2020 but absent in 2018. Conversely, the red arrows correspond 
		to important export transitions existing in 2018 but disappearing in 2020. 
		The solid (dashed) gray arrows are associated to an increase (a decrease) 
		of the corresponding $G_{\rm R}$ matrix element from 2018 to 2020. On the right panels, 
		only the differences between 2020 and 2018
		are highlighted (i.e. only blue and red arrows are kept).
		Countries are represented by their national flags.
	}
\end{figure}

From the reduced Google matrix $\GR$, we easily see the strongest flows of a selected product between the 
countries in 2018. Thus, for {\it food and live animals},
the strongest trade transactions are from: Mexico to USA, Canada to USA,
Japan to China, and Netherlands to Germany. For {\it mineral fuels},
the strongest ones are: Canada to USA, Mexico to USA and the same for {\it machinery}.
Here, we see the domination of USA which has the top PageRank index $K=1$ in 2018
and hence the strongest import flows from certain countries (especially
from its neighbors Mexico and Canada).

The matrix elements of the variation $\Delta \GR$ from 2018 to 2020 have a richer
structure. Thus, for \textit{food and live animals}, the strongest positive variations are observed from UAE to India and to China;
among the negative ones we point variations from Turkey, Russia and India to UAE.
For \textit{mineral fuels}, the strongest positive variations are obtained from Russia and South Korea to China, from Turkey to Netherlands;
among the negative ones we point trade transactions from Japan and South Korea to Australia, from Russia to Netherlands.
Thus, we see that from 2018 to 2020 Russia is changing its \textit{mineral fuels} trade flows by increasing its flow towards China
and decreasing it towards EU represented, here, by the Netherlands.
For \textit{machinery} products, we point among the strongest variations: a positive one from UAE to India,
and a negative one from South Africa to UK.

Summarizing, the REGOMAX approach allows to find the most significant trade flows
between a selected group of countries for a specific product.
The matrix presentation of such trade flows can be represented by a reduced network (Fig.~\ref{fig6}).
Let us consider the reduced Google matrix associated to \textit{food and live animals} (Fig.~\ref{figS3}, top panel). For each column $c$, we select the four largest values corresponding to the four strongest export trade flows from country $c$ to other countries $c'$. This allows to select the most important commercial partners $c'$ importing \textit{food and live animals} products from the country $c$. Hence, from each country of the considered set of interest, ie, here the 20 most important actors of the international trade, we are able to draw four directed links towards 4 other countries. The obtained reduced network highlights the principal \textit{food and live animals} trade interactions between the considered countries. The advantage of the reduced network resides in the fact that, besides direct trade interactions, it displays also long-range trade interactions between the commercial partners. The three panels of Fig.~\ref{fig6} present the reduced networks for \textit{food and live animals}, for \textit{mineral fuels, lubricants and related materials}, and for \textit{machinery, transport equipment} products, respectively.
More precisely, for each type of products, in Fig.~\ref{fig6} left panels, the reduced network for years 2018 and 2020 are jointly displayed: red (blue) links highlight trade interactions disappearing (arising) in 2020 and (not) present in 2018, the solid (dashed) gray links are stable trade interactions the weights of which increase (decrease) from 2018 to 2020. For \textit{food and live animals}, we observe that the main importers are USA, China, Germany, Japan, France and UK. From the top right panel of Fig.~\ref{fig6}, only the changes from 2018 to 2020 are displayed allowing to directly observe the WTN rewiring induced by the COVID-19. From 2018 to 2020, we clearly observe an increase of trade flows towards USA, China and Germany, but a severe decrease towards UK. The latter lose importing flows which benefit in 2020 mainly of China and of Germany.
For \textit{mineral fuels, lubricants and related materials},
the trade flows are polarized towards USA, China, India, Singapore, South Korea, Germany, and the Netherlands. From the middle right panel of Fig.~\ref{fig6}, we observe an increase of trade flows towards Germany and South Korea, whereas Spain loses trade flows to the benefit of South Korea, the Netherlands, and Germany. For \textit{machinery, transport equipment}, importing flows are directed towards USA, Germany, China, France, India, and UK. From the bottom right panel, we observe an increase of the importance of the Netherlands at the expense of Singapore, Japan, and UK. By contrast, UK loses importance to the benefit of the Netherlands and France.

\section{Discussion}
In summary, on the basis of the UN Comtrade database gathering data of the international trade \cite{comtrade}, we contructed the 
multiproduct WTN for the years 2018 to 2020. We
presented its Google matrix analysis based on the PageRank algorithm \cite{brin,meyer} and 
other extended algorithms described in \cite{wtn1,wtn2,wtn3,greduced,politwiki}. 
The PageRank-CheiRank approach allows to obtain a more detailed and advanced 
description of the international trade, in comparison to the usual import-export analysis,
since it takes into account the complex multiplicity of the trade transactions between the countries.
Thus, the PageRank-CheiRank method takes properly into account the robustness and the diversity of the trade connections of a given pair (country, product) highlighting the network influence of
specific countries which is not so visible from the standard import and export description.
As an example, the rankings of the countries with respect to the gas and oil trade are rather different depending on the chosen analysis, either the PageRank-CheiRank analysis or the usual ImportRank-ExportRank analysis.
We point that, in 2018, Russia is at the $K^*=4$ CheiRank position
in contrast to the $\hat{K}^* = 9$ ExportRank position. As the CheiRank takes account of the structure of the WTN, this significant difference
is attributed to the diversity of the trade relations of Russia directed to, both, Asian and European countries.

Using the PageRank-CheiRank algorithms, we determine significant ranking changes of the world countries 
from 2018 to 2020 which are attributed to the COVID-19 impact to the WTN (since there are only
little changes from 2018 to 2019 in absence of the pandemic).
This approach also determines the trade balance $B_c$ of all the world countries in 2018 and 2020, in a manner more adequate than the crude import-export treatment.
The balance variation $\Delta B_c$ between 2018 and 2020 shows that the most 
negatively affected large countries are Ka\-zakh\-stan, Russia, Iran and USA which have been impacted by the global reduction of the world production and related import of {\it mineral fuels}. The PageRank-CheiRank description also allows to determine the sensitivity of the trade balance to the 
price or volume increase of a specific type of products and its variation due to the COVID-19.

New features are also discovered for the trade balance $B_p$ of a product $p$. In contrast to import-export description which, by definition, gives 0 for the balance of each product, our analysis shows
that some products are export oriented:
{\it Food and live animals} ($p=0$),
{\it Beverages and tobacco} ($p=1$), 
{\it Crude materials} ($p=2$),
{\it Mineral fuels} ($p=3$),
{\it Other transactions} ($p=9$);
while others are import oriented:
{\it Animal and vegetable oils} ($p=4$),
{\it Chemicals} ($p=5$),
{\it Manufactured goods} ($p=6$),
{\it Machinery and transport equipment} ($p=7$),
{\it Miscellaneous manufactured articles} ($p=8$). We also determine the variations of these product balances $B_p$
induced by the COVID-19.
Thus, due to COVID-19, the product balance is significantly increased for 
$p=2$ and $3$ products and significantly decreased for $p=6$, $7$, and $8$ products.
We attribute this to 
a significant import reduction of products $p=2$ and $3$, since due to COVID-19,
the production, and hence export, of $p=6$, $7$, and $9$ products were reduced.

Finally, the REGOMAX method, which allows to determine the reduced Google matrix $\GR$ for selected countries and product,
determines the most COVID-affected trade flows and provides a clear graphical
network structure highlighting the rewiring of the WTN induced by the COVID-19 pandemic.

\section{Funding}

This research has been partially supported through the grant
	NANOX N$^\circ$ ANR-17-EURE-0009 (project MTDINA) in the frame 
	of the Programme des Investissements d'Avenir, France. 
	This research has been also supported by the
	Programme Investissements d’Avenir ANR-15-IDEX-0003, 
	ISITE-BFC (GNETWORKS project) and by the Conseil Régional de Bourgogne Franche-Comté (REpTILs project).

\section{Data availability}
The world trade data are available at \url{https://comtrade.un.org}.

\section{Acknowledgements}

We thank the UN Comtrade for a friendly access to their database.

\section{Abbreviations}
	The following abbreviations are used in this manuscript:
	
	\begin{tabular}{@{}ll}
		COVID-19& Coronavirus disease 2019\\
		WTN& World Trade Network\\
		WTO& World Trade Organization\\
		UN& United Nations\\
		WWW& World Wide Web\\
		SITC& Standard International Trade Classification\\
		REGOMAX& Reduced Google matrix\\
		USD& United State Dollar\\
		USA& United State of America\\
		UAE&United Arab Emirates\\
		BRICS& Brazil, Russia, India, China, South Africa\\
		UK& United Kingdom\\
		CC& Country code\\
		EU& European Union
	\end{tabular}

\section{References}

\bibliography{covid}{}

\clearpage
\appendix

\renewcommand\thefigure{\thesection\arabic{figure}}
\setcounter{figure}{0}

\section[\appendixname~\thesection]{Supporting figures}

Here, we present 5 additional figures supporting the results described in the main part of the paper.

\begin{figure}[h]
		\centering
		\includegraphics[width=\textwidth]{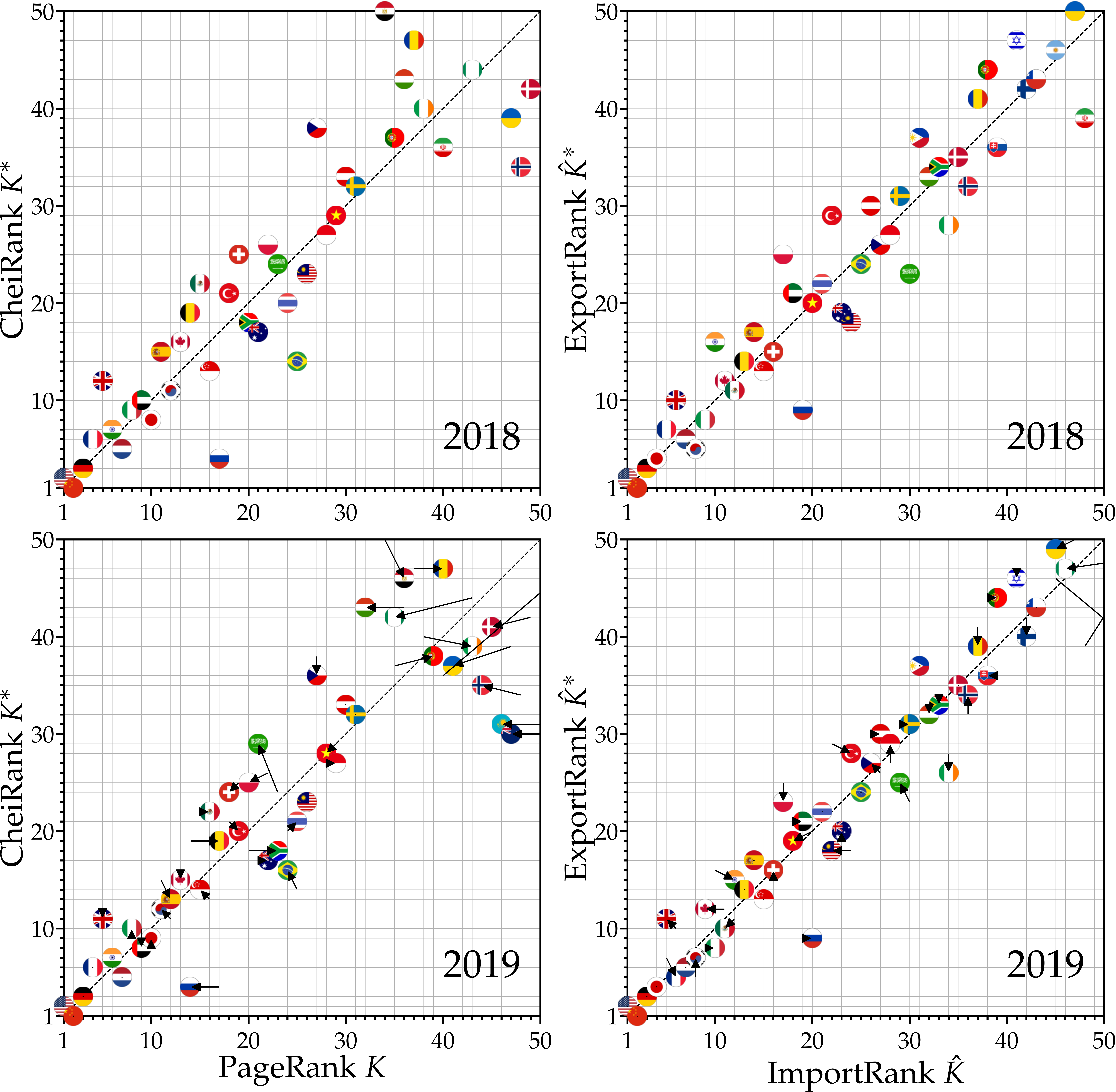}
	\caption{\label{fig1bis}Distributions of countries on 
		the PageRank $K$ - CheiRank $K^*$ plane (left panels) and 
		on the ImportRank $\hat K$ - ExportRank $\hat{K}^*$ (right panels). 
		The distribution are shown for the
		years 2018 (top panels) and 2019 (bottom panels). 
		On the bottom panels, each arrow indicates the change between the two years: 
		the arrow tail [head] is located on the point $\left(K,K^*\right)$ for 2018 [2019]. 
		The countries are represented by their corresponding national flags.}
\end{figure}
\vfill

\vfill
\begin{figure}[t]
		\centering		\includegraphics[width=\textwidth]{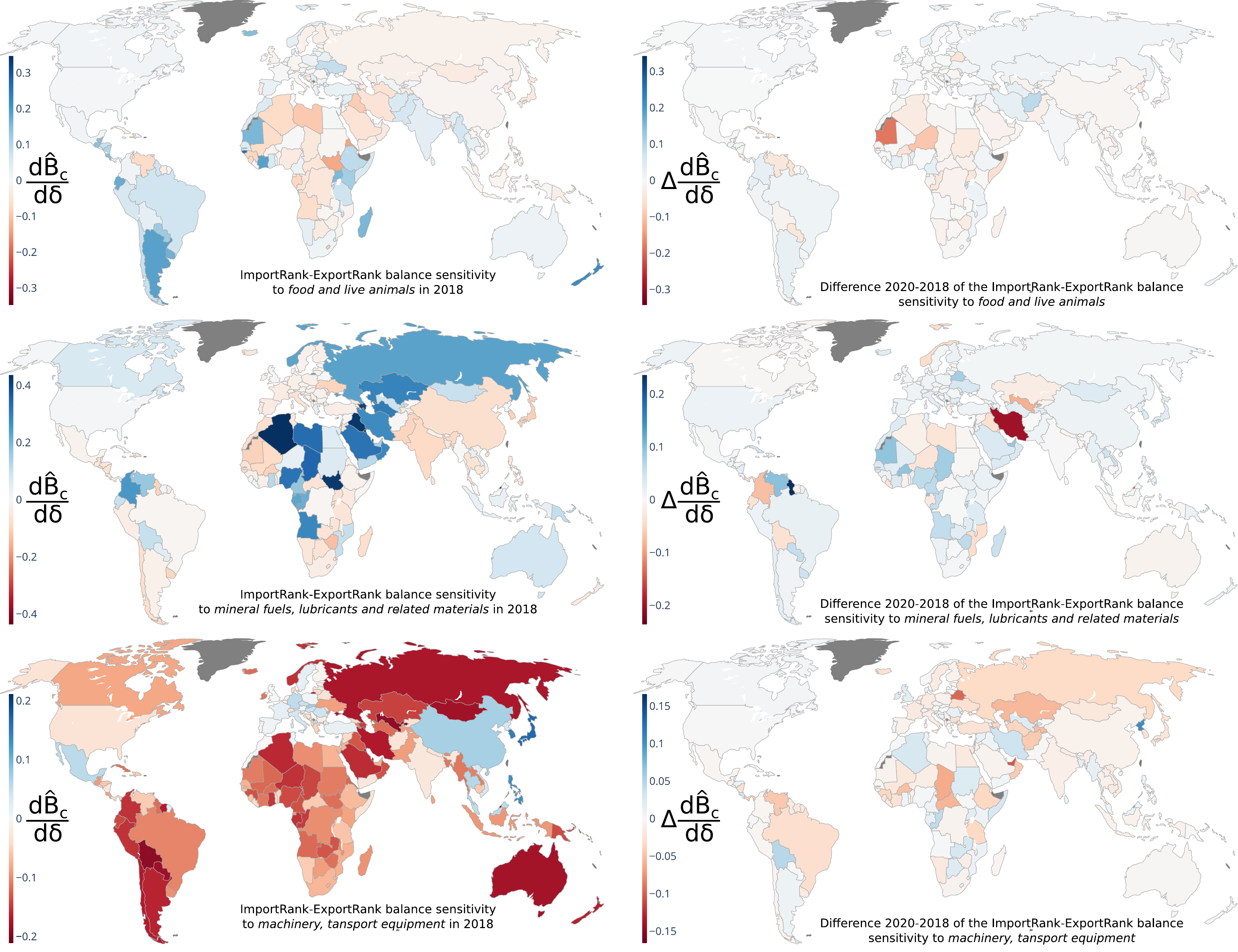}
	\caption{\label{figS1}Geographical distributions of 
		the ImportRank-ExportRank trade balance sensitivity $d\hat{B}_c/d\delta$ to
		\textit{food and live animals} (top row),
		\textit{mineral fuels, lubricants and related materials} (middle row), and
		\textit{machinery, transport equipment} (bottom row) products.
		The left column presents the PageRank-CheiRank trade balance sensitivities for the year 2018.
		The right column presents the difference between the years 2020 and 2018, ie,
		$\Delta d\hat{B}_c/d\delta=\left(d\hat{B}_c/d\delta\right)^{(2020)}-\left(d\hat{B}_c/d\delta\right)^{(2018)}$.
	}
\end{figure}
\vfill

\vfill
\begin{figure}[t]
		\centering		\includegraphics[width=0.75\textwidth]{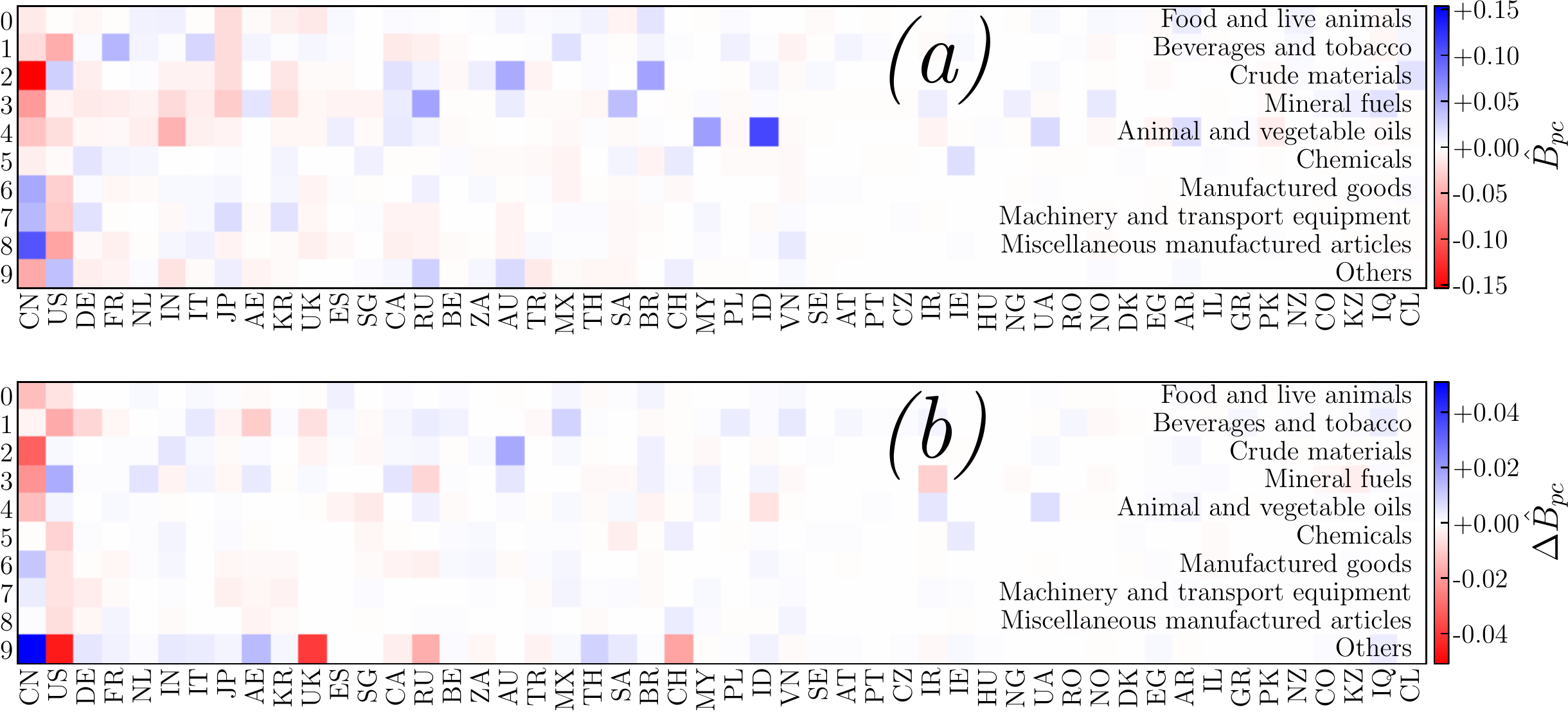}
	\caption{\label{figS2}ImportRank-ExportRank (product,country) trade balance 
		$\hat{B}_{pc}$ [panel $(a)$] for the year 2018.  Trade balance differences between 
		years 2020 and 2018, ie, 
		$\Delta \hat{B}_{pc}=\hat{B}_{pc}^{(2020)}-\hat{B}_{pc}^{(2018)}$, 
		are presented in the panel $(b)$. The columns correspond to countries $c$, sorted according to 2DRank (see Tab.~\ref{tab3} for the first 20),  
		and  rows correspond to products $p$ (see Tab.~\ref{tab1}).
		The meaning of the ISO 3166-1 alpha-2 country codes not present in Tab.~\ref{tab3} can be found, eg, at \cite{wikicc}.
	}
\end{figure}
\vfill

\begin{figure}[t]
\centering
		\includegraphics[width=0.8\textwidth]{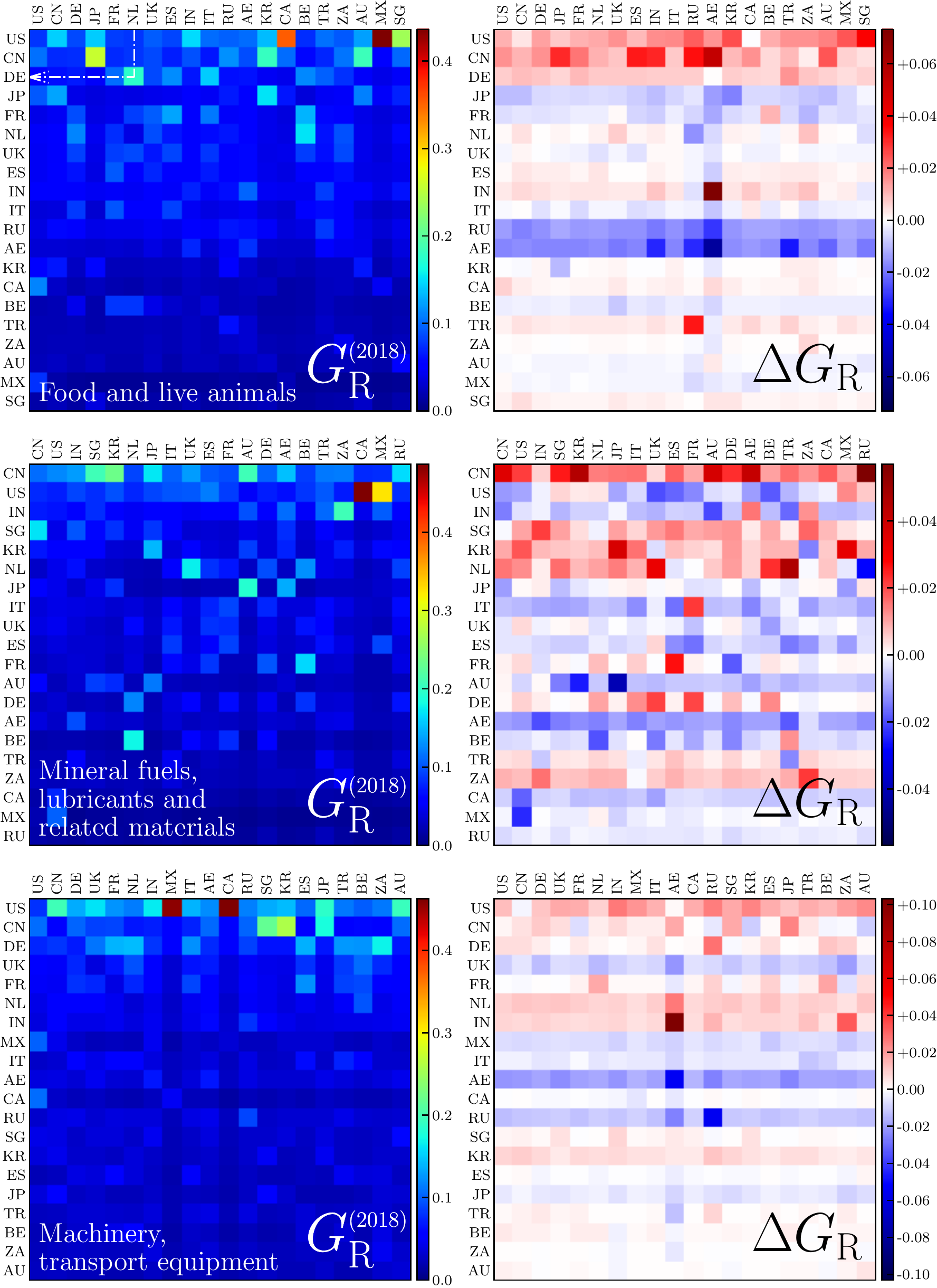}
	\caption{\label{figS3}Reduced Google matrices associated to the
		\textit{food and live animals} (top row),
		\textit{mineral fuels, lubricants and related materials} (middle row), and
		\textit{machinery, transport equipment} (bottom row) products and for the
		20 countries
		at the top of the PageRank and CheiRank lists for the multiproduct WTN (see Tab.~\ref{tab3}).
		The left column presents the reduced Google matrices $\GR^{\left(2018\right)}$ concerning 
		the trade exchanges of the year 2018, and the right column presents 
		the difference of reduced Google matrices between the years 2020 and 2018, ie, 
		$\Delta\GR=\GR^{\left(2020\right)}-\GR^{\left(2018\right)}$.
		The matrix entries correspond to $\left(c,p\right)$ pairs, ie, 
		$\left(c,p=0\right)$ for the top row, $\left(c,p=3\right)$ for the middle row, and 
		$\left(c,p=7\right)$ for the bottom row. These $\left(c,p\right)$-pairs 
		are sorted according to the PageRank algorithm (the country ranks are different for each product $p$). The broken arrow  on the top left panel shows 
		how to read such matrix. The arrow emanates from the country NL entry, goes down 
		and turn left towards the country DE entry. The color at the turning point gives 
		the value, here $\sim0.2$, of the ${\GR}_{\left({\rm DE}\,p=0,{\rm NL}\,p=0\right)}$ matrix element. 
		This matrix element is a measure of the ability of 
		Germany to import \textit{food and live animals} from the Netherlands. 
		By comparison with the other elements of the NL column, this ability 
		is the greatest among the other 19 considered countries. 
		Hence, the most efficient importer from the Netherlands of \textit{food and live animals} is Germany.}
\end{figure}

\vfill
\begin{figure}[t]
		\centering
		\includegraphics[width=0.8\textwidth]{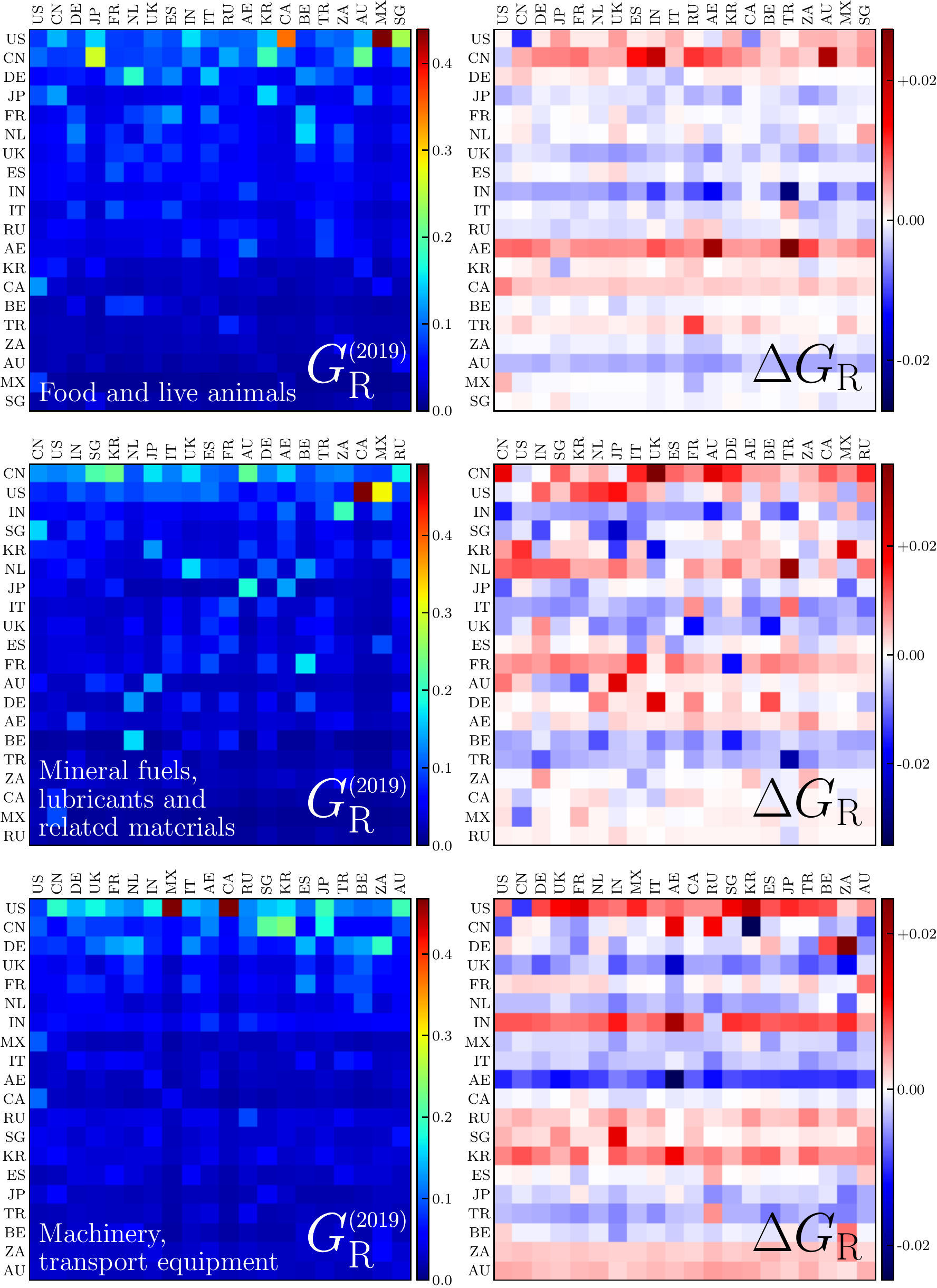}
	\caption{\label{figGR1819}Same as Fig.~\ref{figS3} but for years 2018 and 2019.
	}
\end{figure}
\vfill

\cleardoublepage

\section[\appendixname~\thesection]{Kendall $\tau$ distance}
\label{app:distance}
Let us assume $N$ elements labeled by an integer $\left\{1,\dots,N\right\}$. Let us assume two ranking lists $\tau_1$ and $\tau_2$. The element $i$ occupies the positions $\tau_1(i)$ and $\tau_2(i)$ in these two lists. The Kendall $\tau$ distance between the ranking lists $\tau_1$ and $\tau_2$ is defined as
\begin{equation}
d\left(\tau_1,\tau_2\right)=\displaystyle \left(N\left(N-1\right)\right)^{-1}\sum_{i<j}\left(1-\sign\left(\tau_1\left(i\right)-\tau_1\left(j\right)\right)\sign\left(\tau_2\left(i\right)-\tau_2\left(j\right)\right)\right)
\end{equation}
where the sum is performed over all the different pairs $(i,j)$ of elements and where $\sign\left(x\right)=x/\left|x\right|$.
The Kendall $\tau$ distance $d\left(\tau_1,\tau_2\right)$ ranges from 0 for identical lists, $\tau_1\equiv\tau_2$, to 1 if the $\tau_1$ list is the reverse of the $\tau_2$ list, and \textit{vice versa}.

\end{document}